\documentclass[12pt,preprint]{aastex}

\shorttitle{Probe of Cosmic-Ray Acceleration}
\shortauthors{Fujita et al.}

\begin{document}

\title{Molecular Clouds as a Probe of Cosmic-Ray Acceleration in a
Supernova Remnant}

\author{Yutaka Fujita, Yutaka Ohira,
Shuta J. Tanaka, and Fumio Takahara}
\affil{Department of Earth and Space Science, Graduate School of
Science, Osaka University, 1-1 Machikaneyama-cho, Toyonaka, Osaka
560-0043, Japan}

\begin{abstract}
 We study cosmic-ray acceleration in a supernova remnant (SNR) and the
escape from it. We model nonthermal particle and photon spectra for the
hidden SNR in the open cluster Westerlund~2, and the old-age
mixed-morphology SNR W~28. We assume that the SNR shock propagates in a
low-density cavity, which is created and heated through the activities
of the progenitor stars and/or previous supernova explosions. We
indicate that the diffusion coefficient for cosmic-rays around the SNRs
is less than $\sim 1$\% of that away from them. We compare our
predictions with the gamma-ray spectra of molecular clouds illuminated
by the cosmic-rays (Fermi and H.E.S.S.). We found that the spectral
indices of the particles are $\sim 2.3$. This may be because the
particles were accelerated at the end of the Sedov phase, and because
energy dependent escape and propagation of particles did not much affect
the spectrum.
\end{abstract}

\keywords{radiation mechanisms: non-thermal ---
ISM: individual (\objectname{W~28}) ---
cosmic rays ---
supernova remnants ---
open clusters and associations:
individual (\objectname{Westerlund~2})}

\section{Introduction}

Supernova remnants (SNRs) are canonically considered the main sources of
cosmic-rays in the Galaxy. The detection of non-thermal X-ray emission
from SNRs clearly indicates that electrons are actually accelerated
around the SNR shocks \citep{koy95}, and the observations can constrain
the electron spectra. On the other hand, observational confirmation of
accelerated protons is not as easy as that of electrons. One way to
study the acceleration and spectrum of protons is to study gamma-ray
emission through $pp$-interactions and the decay of neutral pions
\citep*[e.g.][]{nai94,dru94,stu97,gai98,bar99,ber00}. In particular,
molecular clouds are efficient targets of cosmic-ray protons because of
their high density. Thus, clouds illuminated by the protons accelerated
in a nearby SNR could be bright gamma-ray sources
\citep*[e.g.][]{aha96,fat05,fat06,gab09}.

Theoretical studies have suggested that old SNRs could be appropriate
objects to investigate gamma-ray emission through $pp$-interactions,
because the radiation from the accelerated electrons (primary electrons)
disappears as the SNR evolves, owing to their short cooling time
\citep{yam06,fan08}. In other words, we could ignore the gamma-rays from
primary electrons via inverse-Compton (IC) scattering of ambient soft
photon fields and/or non-thermal bremsstrahlung from the interaction of
electrons with dense ambient matter.

In this letter, we consider the evolution of an SNR surrounded by
molecular clouds. We calculate the spectrum of cosmic-rays accelerated
in the SNR and the photon spectrum of the molecular clouds illuminated
by the cosmic-rays. We assume that a supernova explodes in a low-density
cavity, because the progenitor star expels ambient dense gas via strong
UV-radiation and stellar winds \citep{che99}. The cavity may also have
been created through previous supernova explosions. What differentiates
this study is that we consider whether high-energy cosmic-rays
illuminating molecular clouds were accelerated even after the SNR became
old or they were accelerated only when the SNR was young. We also
discuss the influence of particle diffusion on the cosmic-ray
spectrum. We construct specific models for the open cluster Westerlund~2
and the SNR W~28, and compere the results with latest observations.

Westerlund~2 is one of the young open clusters from which TeV gamma-ray
emission has been detected with H.E.S.S. \citep{aha07}. It is surrounded
by molecular clouds \citep{fur09}. The gamma-ray emission is extended
($\sim 0.2^\circ$) and covers the molecular clouds
\citep{fuk09}. Noticeable objects such as pulsars that can be the source
of the gamma-ray emission have not been observed in this region.
\citet{fuj09a} proposed that the gamma-ray emission comes from an old
SNR, although there is no clear signature of SNRs in the cluster.  W~28
is a mixed-morphology SNR interacting with a molecular cloud
\citep{woo81}. It is an old SNR and TeV gamma-rays have been detected
from molecular clouds around the SNR \citep{aha08b}.

\section{Diffusion of Cosmic-Ray Particles}

As will be shown in the next section, the proton spectrum around
Westerlund~2 and W~28 can be fitted with a power-law with an index
$s\sim 2.3$ (the definition is shown in
equation~[\ref{eq:NpE}]). Moreover, the shock waves seem to have
finished accelerating particles for those objects, while the surrounding
regions are bright in the gamma-ray band. Here, we briefly discuss what
they mean before we explain assumptions in our specific models for
Westerlund~2 and W~28.

\subsection{Energy Spectrum of Cosmic-Rays}

If the duration of particle acceleration is shorter than that of the
diffusion, and the particle source is spatially localized well, we can
use the analytical solution in \citet*{ato95}. This corresponds to the
situation where particles are accelerated mainly when the SNR is young
and compact, and the molecular cloud illuminated by accelerated
cosmic-rays is distant from the SNR shock. If the shape of the source
spectrum is a power-law with an index $\alpha$ or $Q\propto
E^{-\alpha}\delta(r)\delta(t)$, the energy spectrum at the position of
the cloud ($r$) is represented by
\begin{equation}
 f(E) \propto \frac{E^{-\alpha}}{r_{\rm
  diff}^3}\exp\left(-\frac{r^2}{r_{\rm diff}^2}\right)\:,
\end{equation}
if radiative cooling during the diffusion can be ignored. The diffusion
length is $r_{\rm diff}=2\sqrt{D(E)t}$, where $D(E)$ is the diffusion
coefficient. Following \citet{gab09}, we assume that
\begin{equation}
\label{eq:DE}
 D(E)=10^{28}\chi \left(\frac{E}{10\rm\: GeV}\right)^{0.5}
\left(\frac{B}{3\:\mu\rm G}\right)^{-0.5}\rm\: cm^2\: s^{-1}\:,
\end{equation}
where $B$ is the magnetic field. At a point distant from an SNR, we
expect that $\chi\sim 1$. Thus, for a given magnetic field, the energy
spectrum is represented by $f\propto E^{-\alpha-0.75}$ if $r\lesssim
r_{\rm diff}$ at the position of the molecular cloud. This means that
even if particles are accelerated efficiently ($\alpha=2$), the energy
spectrum must be soft ($f\propto E^{-2.75}$). In other words, if the
index of the spectrum is observed to be $s<2.75$ at a molecular cloud,
it is likely that the particles are accelerated near the cloud after the
SNR becomes large.

For Westerlund~2 and W~28, since the spectral indices are $s\sim 2.3$
for the high-energy protons ($\gtrsim$~TeV) that are illuminating the
molecular clouds around these objects, we expect that the cosmic-rays
were accelerated near the molecular clouds even after the SNRs became
old and large. This may be in contrast with the assumption often adopted
in theoretical studies. We assume that the SNR shock had traveled in a
low-density cavity. During the propagation in the cavity, the shock wave
is in the adiabatic Sedov phase, because the low-density prevents the
radiative cooling. Thus, even if particles can be accelerated only
during the Sedov phase, they are being accelerated until the shock
reaches and collides with the surrounding high-density region, which is
an effective target of the cosmic-rays. The particles illuminate the
high-density region with the energy spectrum at the acceleration site or
the shock. Thus, the spectral indices of $s<2.75$ are possible.

\subsection{Diffusion Coefficient}

For Westerlund~2 and W~28, efficient particle acceleration around shock
waves seems to have finished. For the former, there is no signature of a
shock. Thus, the shock may have already collided with the surrounding
dense region and dissipated. For the latter, the shock is traveling in a
relatively high density region ($\sim 10$--$100\rm\: cm^{-3}$) and seems
to be in a radiative phase \citep{rho02}. Thus, cosmic-rays that are
illuminating molecular clouds might be accelerated in the past. For
$\chi =1$ and $B=20\rm\: \mu G$ in equation (\ref{eq:DE}), \citet{gab09}
estimated the diffusion time of cosmic-rays in a typical molecular cloud
and showed that it is only $\sim 100$~yr for protons with an energy of
$\sim 10$~TeV. However, the probability that the particle acceleration
has finished within $\sim 100$~yr must be very small. Therefore, in the
next section we simply assume that $\chi\ll 1$ around the SNRs and later
discuss it quantitatively. The small $\chi$ may be because of the
generation of plasma waves by cosmic-rays \citep{wen74}.

Note that if particles are accelerated with a spectral index of
$\alpha=2$, and if they diffuse in the interstellar gas and illuminate
molecular clouds before the index increases to $s=2.75$, the observed
spectral index could be $s\sim 2.3$ as the ones observed in Westerlund~2
and W~28. However, the value of $\chi$ and/or the distance between the
SNR and the molecular clouds must be fine-tuned in this case. In
particular, for W~28, the molecular clouds are adjacent to the SNR
\citep{aha08b} and there seems to be no room to adjust the distance.
Thus, we do not consider this possibility.

\section{Models and Results}

\subsection{Westerlund 2}

We assume that the distance to Westerlund~2 is $d=5.4$~kpc
\citep{fur09}. We also assume that a supernova exploded at the center of
a cavity, which is filled with warm gas, and is surrounded by molecular
clouds (Fig.~\ref{fig:conf}a). After the explosion, a shock expands in
the warm gas and then hits the surrounding molecular gas ($\sim 20$~pc
from the cavity center). The size of the cavity is based on the arc
structure found by \citet{fuk09}.

When the shock expands in the cavity, the shock velocity $v_s$ is
written as a function of the SNR age ($t$) as
\begin{equation}
v_s(t)=
\left\{
\begin{array}
{l@{} l@{}}
v_i & (t<t_1) \\
v_i (t/t_1)^{-3/5} ~
& (t_1<t<t_2) 
\end{array} \right.~~ ,
\label{eq:Vs}
\end{equation}
where $v_i=v_{i,9}10^9$~cm~s$^{-1}$ is the initial velocity of the
ejecta \citep{stu97,yam06,fan08}. At $t<t_1 =2.1\times 10^2\:
(E_{51}/n_0)^{1/3}v_{i,9}^{-5/3}$~yr, the SNR is in the free expansion
phase, at $t>t_2=4.0\times 10^4 E_{51}^{4/17}n_{\rm 2}^{-9/17}$~yr, it
is in the radiative phase, and at $t_1<t<t_2$, it is in the Sedov
phase. Here $n_0= 100\: n_2\:\rm cm^{-3}$ and $n_0= \mu\: n_{\rm
cav}\:\rm cm^{-3}$, where $n_{\rm cav}$ is the hydrogen density in the
cavity and $\mu=1.4$ is the mean atomic weight of the gas assuming one
helium atom for every 10 hydrogen atoms. The initial energy of the
ejecta is $E_{\rm SN}=E_{51}10^{51}$~erg. The radius of the shock
($r_s$) can be obtained by integrating equation~(\ref{eq:Vs}).

The energy spectrum of the accelerated protons is given by
\begin{equation}
\label{eq:NpE}
 N_p(E)\propto E^{-s}\exp(-E/E_{\rm max, p})\; .
\end{equation}
The index is given by $s=(r_{\rm com}+2)/(r_{\rm com}-1)$, where $r_{\rm
com}$ is the compression ratio of the shock \citep{bla87}, which is
given by the Rankine-Hugoniot relation for $t_1<t<t_2$. We do not
consider non-linear effects, and thus $r_{\rm com}$ is equal to or
smaller than 4. We assume that the maximum energy is determined by the
age of the SNR:
\begin{equation}
 E_{\rm max,p}=1.6\times10^2~h^{-1} 
\left(\frac{v_s}{10^8\rm\: cm\: s^{-1}}\right)^2
\left(\frac{B_{\rm d}}{10~\rm\mu G}\right)
\left(\frac{t}{10^5{\rm yr}}\right)~{\rm TeV}~,
\label{eq:Emax_p}
\end{equation}
where $h~(\sim 1)$ is the factor determined by the shock angle and the
gyro-factor, and $B_{\rm d}$ is the downstream magnetic field, which is
given by $B_{\rm d}=r_{\rm com}B_{\rm cav}$, where $B_{\rm cav}$ is the
magnetic field in the cavity \citep{yam06}. The minimum energy of the
protons is given by the rest-mass energy.

In our fiducial model, we assume that $n_{\rm cav}=4\rm\: cm^{-3}$,
which is the lower end of the typical value in such regions ($\sim
5$--$25/\mu\rm\: cm^{-3}$; \citealt{che99}). There is no information
about the temperature and magnetic field in the cavity before the
explosion. Thus, we assume that $T_{\rm cav}=8\times 10^5$~K and $B_{\rm
cav}=20\:\mu\rm G$ so that the photon spectrum is consistent with
observations; $T_{\rm cav}$ and $B_{\rm cav}$ determine $s$ and $E_{\rm
max,p}$, respectively. It is to be noted that the current temperature of
the warm gas is $1.4\times 10^6$--$4.9\times 10^7$~K \citep{fuj09a}. For
$v_{i,9}=E_{51}=1$, the shock radius is $r_s=20$~pc at $t=1.6\times
10^4$~yr. We define this time as $t_{\rm coll}$ and the shock collides
with the molecular clouds around this time. Since $t_2=1.6\times
10^4$~yr, the SNR is at the end of the Sedov phase at $t=t_{\rm
coll}$. Assuming that particle acceleration stops at $t\sim t_2$, the
protons accelerated at $t\lesssim t_{\rm coll}$ ($\sim t_2$) illuminate
the molecular clouds. For $t_1<t<t_{\rm coll}$, the compression ratio
$r_{\rm com}$ and the proton cut-off energy $E_{\rm max,p}$ decrease
from 4 to 3.6 and from 151 to 47~TeV, respectively. Therefore, the
protons that illuminate the molecular clouds have a spectrum represented
by equation~(\ref{eq:NpE}) with $2.0\leq s\leq 2.3$ and $47\leq E_{\rm
max,p}\leq 151$~TeV.

The mass of the molecular clouds around Westerlund~2 is $\sim 2\times
10^5\rm M_\sun$ \citep{fur09}, although the clouds are very
irregular. Referring to the observations, we assume that the proton
density of the clouds is $n_c=1.0\times 10^3\rm\: cm^{-3}$. In the
clouds, we expect that the shock follows the ``snowplow'' evolution
($t>t_{\rm coll}$). We assume that the shock in the clouds can locally
be expressed as a shell centered on $r=0$. In this case, from the
momentum conservation, the radius of the shell, $r_{\rm sh}(t)$, has a
relation of
\begin{equation}
\label{eq:mom} \frac{4\pi}{3}\left[n_c (r_{\rm sh}(t)^3-r_s(t_{\rm
 coll})^3) + n_{\rm cav} r_s(t_{\rm coll})^3\right]\dot{r}_{\rm sh}(t) =
 \frac{4\pi}{3}n_{\rm cav} r_s(t_{\rm coll})^3 v_s(t_{\rm coll}) \:,
\end{equation}
with $r_{\rm sh}=r_s$ at $t=t_{\rm coll}$, which can be solved
numerically for $t>t_{\rm coll}$. After the collision, it takes $t_{\rm
stop}\sim 1.8\times 10^4$~yr until the shell velocity decreases to the
internal velocity of the clouds ($\sim 20\rm\: km\: s^{-1}$;
\citealt{fur09}). Since the shock has not been observed apparently, its
velocity may have decreased to that level.

Some of the accelerated protons plunge into the molecular clouds and
create electrons, positrons, and gamma-ray photons through
$pp$-interactions there. The electrons and positrons (both are called
secondary electrons) also emit photons through synchrotron, IC, and
bremsstrahlung radiation. Figure~\ref{fig:wd2} shows the broad-band
spectrum of the molecular clouds at $t\sim t_{\rm coll}$. We assume that
the spectrum of the cosmic-ray protons is represented by
equation~(\ref{eq:NpE}) with $s=2.3$ and $E_{\rm max,p}=47$~TeV. The
total energy of the protons illuminating the molecular clouds is
$4.5\times 10^{49}$~erg. Note that the total number of the target
protons and the total energy of the illuminating protons are degenerated
for the photon spectrum. The magnetic field in the cloud is $B_{\rm
c}=60\:\mu\rm G$. The spectrum is calculated based on the radiation
models in \citet{fan08}. For the $pp$-interactions, we used the code
provided by \citet{kal08}. In Figure~\ref{fig:wd2}, we included the
contribution of background cosmic-rays (equation~[10] in
\citealt{gab09}), but it is negligibly small. We assumed that the
injection of secondary electrons is balanced with the cooling. The model
results are compared with observations. The radio flux is considered as
an upper limit, because it could only partially include the non-thermal
radio flux produced by the energetic electrons \citep{whi97}. For the
two-band Fermi observations, we assume that the spectral index is
between $-2.5$ and $-2$. The data were particularly important to
constrain the value of $s$.

The diffusion time of cosmic-rays in a cloud is written as
\begin{equation}
 t_{\rm diff}\sim \frac{L_c^2}{6 D(E)}
=1.6\times 10^4\left(\frac{\chi}{0.01}\right)^{-1}
\left(\frac{L_c}{15\rm\: pc}\right)^2
\left(\frac{E}{10\rm\: TeV}\right)^{-0.5}
\left(\frac{B_c}{60\rm\: \mu G}\right)^{0.5} \rm yr
\:,
\end{equation}
where $L_c$ is the size of a molecular cloud.  From the condition of
$t_{\rm diff}\gtrsim t_{\rm stop}$, the reduction factor must be
$\chi\lesssim 0.01$. If we ignore the diffusion of the protons in the
cloud, the emission originated from the protons will not change for a
long time because of the long cooling time of the protons ($>10^5$~yr;
\citealt{gab09}).

In Figure~\ref{fig:wd2}, the emission from primary electrons is also
presented. Their energy spectrum is given by
\begin{equation}
\label{eq:NeE}
 N_e(E)\propto E^{-s}\exp(-E/E_{\rm max, e})\; .
\end{equation}
The maximum energy is determined by synchrotron cooling:
\begin{equation}
 E_{\rm max,e}=14~h^{-1/2} v_{s,8}
(B_{\rm d}/10~\rm\mu G)^{-1/2}~{\rm TeV}~,
\label{eq:Emax_e}
\end{equation}
\citep{yam06}. The minimum energy of the electrons is given by the
rest-mass energy. We assume that the electron-proton ratio is $K_{\rm
ep}=0.01$. If the bremsstrahlung radiation from primary electrons were
more than ten times brighter than that in Figure~\ref{fig:wd2}, it might
overwhelm the gamma-ray emission from the pion decay. However, the
required electron-proton ratio ($K_{\rm ep}\gtrsim 0.1$) would be
unrealistically large. Moreover, because of the short cooling time, the
radiation of the primary electrons should decrease after particle
acceleration stops. Thus, for $t>t_{\rm coll}$ ($\sim t_2$), observed
flux from the primary electrons should be lower than that in
Figure~\ref{fig:wd2}.

\subsection{W 28}

\begin{figure}
\epsscale{.80} \plotone{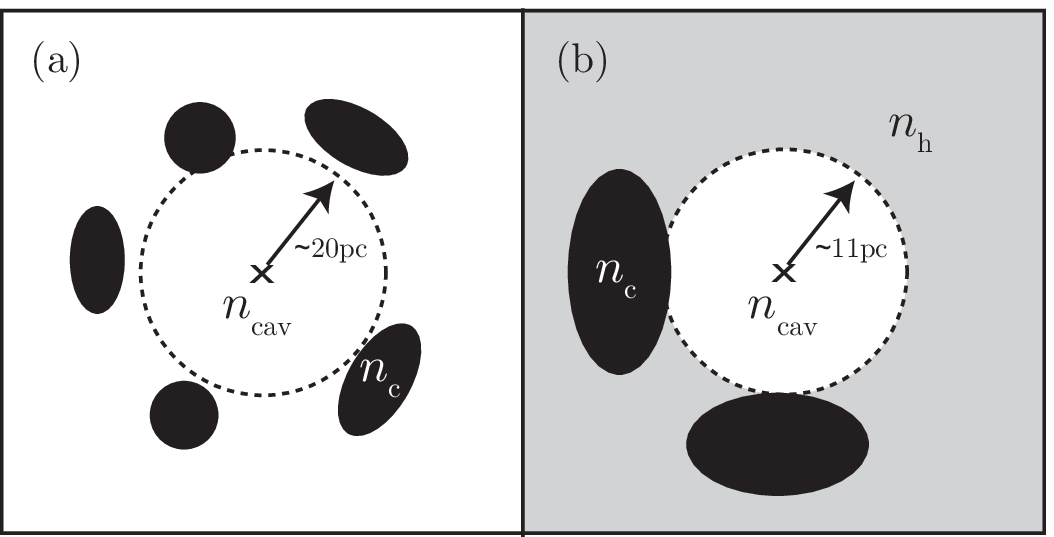} \caption{Configuration of the cavity and
clouds. (a) Westerlund~2 and (b) W~28. \label{fig:conf}}
\end{figure}

We apply the same model to W~28. A supernova explodes in a low-density
cavity (Fig.~\ref{fig:conf}b). The distance to W~28 has been estimated
to be $d\sim 2$--3~kpc \citep{gou76,loz81}; we assume $d=2.6$~kpc in
this letter. We assume that $n_{\rm cav}=8\rm\: cm^{-3}$, $B_{\rm
cav}=2\:\mu\rm G$, and $T_{\rm cav}=1\times 10^6$~K. For $v_{i,9}=1$ and
$E_{51}=0.4$ \citep{rho02}, the shock radius is $r_s=11$~pc at
$t=t_2=9.0\times 10^3$~yr. During the Sedov phase, the index $s$
increases from 2.0 to 2.3, and the maximum proton energy $E_{\rm max,
p}$ decreases from 8.8 to 2.7~TeV. At $t\sim t_2$, we assume that the
shock collides with a surrounding relatively dense region of
$n_h=70\rm\: cm^{-3}$ (Fig.~\ref{fig:w28}b) and we define this time as
$t=t_{\rm coll}$. The density $n_h$ is based on observations
\citep{lon91,rea00}. The detection of 1720 MHz maser emission from W~28
actually indicates the presence of the shock in the dense gas
\citep{fra94,cla97}. After the collision, the evolution of the shock
follows equation~(\ref{eq:mom}) in which $n_c$ is replaced by $n_h$. The
radius of the shell increases from $r_{\rm sh}=11$ to 13~pc, and the
velocity decreases from $v_s=500$ to $80\rm\: km\: s^{-1}$ in $t_{\rm
stop}\sim 1.4\times 10^4$~yr. The radius of $r_s=13$~pc and the velocity
of $80\rm\: km\: s^{-1}$ are comparable to the observed ones
\citep{boh83,rho02}. Note that even if particles can be accelerated at
radiative shocks, it cannot be applied to W~28 at present, because the
small velocity ($80\rm\: km\: s^{-1}$) prevents ionization of the
preshock gas \citep{shu79}.

Since the gamma-ray emission comes from clumps around the SNR
\citep{aha08b}, we expect that there are dense clumps, which are the
targets of cosmic-ray protons (Fig.~\ref{fig:conf}b). Referring to the
observations, we assume that the mass and density of a clump is $4\times
10^4\: M_\sun$ and $n_c=5.4\times 10^2\rm\: cm^{-3}$, respectively
\citep{aha08b}.  Figure~\ref{fig:w28} shows the broad-band spectrum of
the molecular cloud at $t\sim t_{\rm coll}$. The spectrum of the
cosmic-ray protons is represented by equation~(\ref{eq:NpE}) with
$s=2.3$ and $E_{\rm max,p}=2.7$~TeV. The total energy of the protons
that contribute to the illumination of the molecular clouds is $2\times
10^{49}$~erg. The magnetic field in the cloud is $B_{\rm c}=60\:\mu\rm
G$. The radio flux is considered as an upper limit, because it includes
radio emission outside of the region where TeV gamma-ray has been
detected \citep{dub00}. Since the XMM-Newton upper-limit is for part of
the TeV gamma-ray region, the actual upper-limit may be larger (Nakamura
et al. 2009, in preparation). As is the case for Westerlund~2, while the
photon flux from primary electrons decreases as the time advances, that
originated from protons does not. The condition $t_{\rm diff}\gtrsim
t_{\rm stop}$ gives $\chi \lesssim 0.01$. Recently, Fermi found GeV
gamma-rays in the molecular clouds where TeV gamma-ray has been detected
\citep{abd09}, which indicates that the energy dependence of $D(E)$ is
not affecting the proton spectrum and is consistent with our assumption
that the protons were accelerated near the molecular clouds.

\section{Summary and Discussion}

\begin{figure}
\epsscale{.60} \plotone{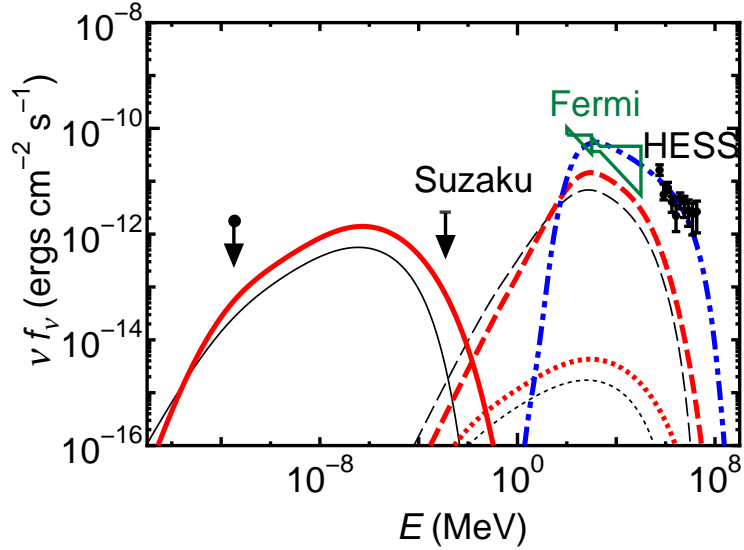} \caption{Comparison of the model results
with radio \citep{whi97}, Suzaku \citep{fuj09a}, Fermi \citep{abd09},
and H.E.S.S. observations \citep{aha07} for Westerlund~2. The
synchrotron radiation (solid line), bremsstrahlung (dashed line), and IC
scattering (dotted line) are of the primary electrons (thin lines) and
the secondary electrons (thick lines). The $\pi^0$ decay gamma-rays are
shown by the two-dot-dashed line. \label{fig:wd2}}
\end{figure}

\begin{figure}
\epsscale{.60} \plotone{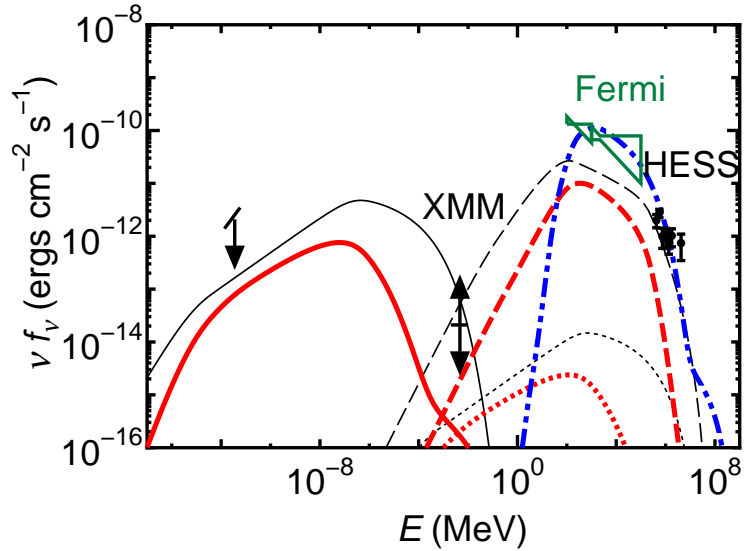} \caption{Comparison of the model results
with radio \citep{dub00}, XMM-Newton (Nakamura et al. 2009, in
preparation), and Fermi observations \citep{abd09} for
W~28. H.E.S.S. observations are for HESS J1801-233 \citep{aha08b}. The
lines are the same as those in Fig.~\ref{fig:wd2}.\label{fig:w28}}
\end{figure}

In this letter, we constructed a model of gamma-ray emission from
molecular clouds surrounding an SNR. We considered old SNRs because
gamma-rays mainly come through $pp$-interactions and pion decay, and
those from primary electrons can be ignored. In fact, the synchrotron
cooling time of electrons with $E\sim 1$~TeV is only $\sim 4\times
10^3$~yr for $B\sim 60\:\rm \mu G$ \citep{gab09}, and thus they cool
quickly after their acceleration finishes. We assumed that the SNR shock
travels in a low-density cavity, and we indicated that the diffusion
coefficient of cosmic-ray particles around the SNR needs to be much
smaller than that away from it (less than $\sim 1$\%), The predicted
broad-band spectra for Westerlund~2 and W~28 are consistent with
observations (Figs.~\ref{fig:wd2} and~\ref{fig:w28}). The proton spectra
can be fitted with a power law with indices $\sim 2.3$, which is the
value at cosmic-ray sources expected in cosmic-ray propagation models
for the Galaxy \citep{str07}. The indices suggest that particles were
accelerated near the molecular clouds even after the SNR became old and
large for these objects. The soft spectra may be because the particles
were accelerated at the end of the Sedov phase. While the low Mach
number of the shock makes the spectra soft in our model, neutral
hydrogens in the cavity may also do that \citep*{ohi09}. On the other
hand, for another SNR IC~443, the broad-band photon spectrum indicates
that the index is 2 \citep{zha08}. For this object, protons accelerated
at an earlier stage of the Sedov phase (when the Mach number of the
shock is large) may be illuminating molecular clouds. Moreover, if
molecular clouds are distributed with various sizes and distances from
the SNR, the gamma-ray spectrum may be the superposition of the spectrum
of each cloud and may reflect the energy-dependence of particle
diffusion (equation~[\ref{eq:DE}]; see \citealt{aha96,tor08}).

\acknowledgments

We are grateful to Y.~Fukui, R.~Nakamura, A.~Bamba, R.~Yamazaki, and
K.~Kohri for useful discussion. This work was supported by KAKENHI
(Y.~F.: 20540269; Y.~O.: 20.1697; F.~T.; 20540231).

\clearpage

\clearpage

\end{document}